\documentclass[aps,pra,floatfix,showpacs,groupedaddress,notitlepage,twocolumn,amsmath,amssymb]{revtex4}

\usepackage{graphicx}
\usepackage{bm}
\usepackage{braket}
\usepackage{times} 
\expandafter\let\csname equation*\endcsname\relax
\expandafter\let\csname endequation*\endcsname\relax
\usepackage{amsmath}
\usepackage{amssymb}
\usepackage{amstext}
\usepackage{color}
\usepackage{units}

\newcommand{\fig}[1]{Fig.~\ref{#1}}
\renewcommand{\sec}[1]{Sec.~\ref{#1}}
\renewcommand{\quote}[1]{Ref.~\cite{#1}}

\newcommand{\ph}{\mathrm{ph}}

\begin{document}

\title{Dimerized Mott insulators in hexagonal optical lattices} 
\author{Ole J\"urgensen,$^1$ and Dirk-S\"oren L\"uhmann,$^1$}
\affiliation{$^1$  Institut f\"ur Laser-Physik, Universit\"at Hamburg, Luruper Chaussee 149, 22761 Hamburg, Germany}
 

\begin{abstract}

We study bosonic atoms in optical honeycomb lattices with anisotropic tunneling and find dimerized Mott insulator phases with fractional filling. These incompressible insulating phases are characterized by an interaction-driven localization of particles in respect to the individual dimers and large local particle-number fluctuations within the dimers. We calculate the ground-state phase diagrams and the excitation spectra using an accurate cluster mean-field method. 
The cluster treatment enables us to probe the fundamental excitations of the dimerized Mott insulator where the excitation gap is dominated by the intra-dimer tunneling amplitude. This allows the distinction from normal Mott insulating phases gapped by the on-site interaction.
In addition, we present analytical results for the phase diagram derived by a higher-order strong-coupling perturbative expansion approach. By computing finite lattices with large diameters the influence of a harmonic confinement is discussed in detail. It is shown that a large fraction of atoms forms the dimerized Mott insulator under experimental conditions. The necessary anisotropic tunneling can be realized either by periodic driving of the optical lattice or by engineering directly a dimerized lattice potential. The dimers can be mapped to to their antisymmetric states creating a lattice with coupled $p$-orbitals.  
\end{abstract}


\maketitle


In recent years, experiments with ultracold atoms in optical lattices have attracted a lot of attention due to their unique possibilities to simulate condensed matter systems in a highly controllable environment. The pioneering experiments in this field have been realized in cubic lattices demonstrating the observability 
 of the transition between the superfluid and the Mott insulating phase \cite{Jaksch1998,Greiner2002}. Lately, more sophisticated setups allow experiments with a variety of non-cubic optical lattice geometries including superlattices \cite{Guidoni1997, Peil2003, Santos2004, Anderlini2007, Trotzky2008}, checkerboard \cite{Sebby-Strabley2006, Wirth2011}, Kagom\'e \cite{Jo2012},  and honeycomb lattices \cite{Becker2010,SoltanPanahi2011, SoltanPanahi2012,Tarruell2012,Greif2013,Uehlinger2013,Luhmann2014}. The latter is of particular interest due to its analogy to graphene \cite{Zhu2007,Wu2008,Chen2011,Zhang2012,Polini2013}. Furthermore, the periodic modulation of the phase of the lattice beams offers an additional tool to engineer the tunneling matrix elements \cite{Eckardt2005,Lignier2007,Struck2011,Struck2012,Hauke2012}. 

\begin{figure}[b]
\includegraphics[width=\linewidth]{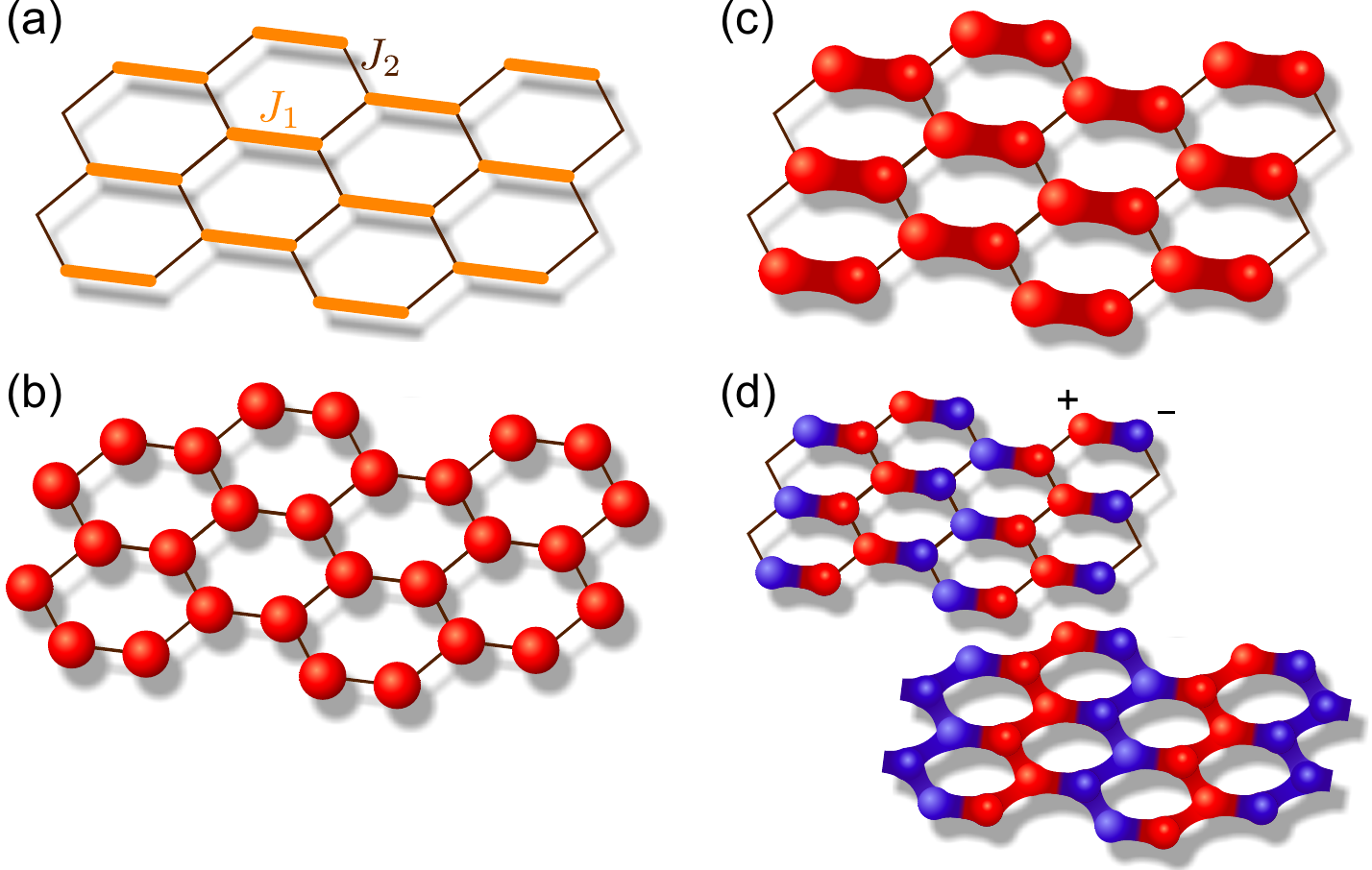} \\[-1em] \centering
\caption{
(a) A honeycomb lattice with anisotropic tunneling matrix elements $J_1$ and $J_2$. The sites within a dimer are coupled by the tunneling matrix element $J_1$, while the coupling between neighboring dimers is associated with the smaller tunneling $J_2$. For $J_2\ll J_1$ a quasi-quadratic lattice structure of dimers is formed. 
(b) In the Mott insulator phase with integer filling $\rho=1$ each atom is localized at a lattice site.
(c) In the \textit{dimerized} Mott insulator phase with $\rho=1/2$, $3/2$, $...$ the atoms are still delocalized within individual dimers, while the superfluid order parameter of the lattice vanishes.
(d) By lattice shaking the tunneling matrix elements along the dimers can be tuned to negative values, where each dimer wave function resembles a $p$-orbital state. In the superfluid state a stripe order is formed.}
\label{fig:lattice}
\end{figure}

Superlattices with different site offsets and barrier heights can be created by superposing lattices with different wavelengths \cite{Roth2003, Peil2003, Rabl2003, Santos2004, Buonsante2004a, Sebby-Strabley2006,Rousseau2006, Anderlini2007, Trotzky2008, Tarruell2012,Uehlinger2013,Greif2013} or by employing polarization-dependent light potentials \cite{Becker2010,SoltanPanahi2011, SoltanPanahi2012, Luhmann2014}. The unit cells of these lattices consist of multiple sites, adding a new degree of complexity to the system. If the sites within the unit cell have different energy offsets, normal Mott insulators with a population imbalance emerge \cite{Roth2003,Rabl2003,Buonsante2004a,Chen2010,Luhmann2014}, where the atom are localized on individual sites. For the case without site-offsets, anisotropic tunneling couplings lead to non-trivial insulator phases with fractional filling in between the conventional Mott insulator phases \cite{Buonsante2004b, Buonsante2005a, Buonsante2005b, Danshita2007, Muth2008}. Here, the particles localize on individual unit cells with a vanishing superfluid order parameter but are still delocalized within each unit cell. Originally, respective experiments have been proposed for one-dimensional superlattices, but so far, these phases with fractional filling have not been observed. For this the reason is two-fold. First, the phase diagram suggests a very small fraction of atoms to occupy the dimerized phase in a confined system. Second, a clear signature for discriminating the fractionally filled phase from the conventional Mott insulator was missing.

 In this work, we  theoretically study bosonic atoms in honeycomb optical lattices with adjustable tunneling matrix elements. The latter is achieved either by a periodic driving of a honeycomb lattice or by engineering directly a dimerized potential \cite{Tarruell2012,Greif2013,Uehlinger2013}. In addition to normal Mott insulating phases with integer filling, we find insulating phases with half-integer filling where the particles are delocalized on dimers. The dimers are naturally defined by the biatomic unit cell of the honeycomb lattice, allowing for non-trivial fractional-filling phases \cite{Buonsante2004b, Buonsante2005a, Buonsante2005b, Danshita2007, Muth2008}. The phase diagram is studied by means of the strong-coupling expansion approach similar to \quote{Buonsante2005a} as well as by the cluster Gutzwiller mean-field method giving accurate results for honeycomb lattices \cite{Luhmann2013}. 
This method grants the great advantage of the access to the excitation spectrum. We show that the characteristic local excitations allow distinguishing experimentally between the conventional and this \textit{dimerized Mott insulator} state. The excitation gap gives an estimate of the required temperatures for observing this quantum phase. Furthermore, we simulate two-dimensional lattice planes with harmonic confinement using realistic experimental parameters. We find that the dimerized state is formed by a large fraction of atoms ($> 70\%$) and thus is well observable in the proposed experiment.

\section{Phase diagram}

There are two perspectives to realize the dimerized Mott insulator phase in experiments with hexagonal lattices. 
First, we propose to use an optical honeycomb lattice generated by three running laser beams as in Refs.~\cite{Becker2010,SoltanPanahi2011, SoltanPanahi2012}. A honeycomb lattice features a two-atomic unit cell, where the intra-cell bond and inter-cell bonds have different orientations, which allows addressing them independently by lattice shaking techniques. 
By modulating the relative phases of the beams, the lattice is periodically accelerated on an elliptical orbit \cite{Eckardt2005,Struck2011}. Employing Floquet theory, one can obtain an effective time-averaged Hamiltonian, where the tunneling matrix elements are modified by a Bessel function depending on the driving parameters \cite{Eckardt2005,Lignier2007,Struck2011,Struck2012,Hauke2012}.
This allows engineering two different and tunable tunneling matrix elements $J_1$ and $J_2$  in the vertical and horizontal direction (see \fig{fig:lattice}a). When $J_1$ is larger than $J_2$, the honeycomb lattice separates into a dimerized square lattice of double-wells coupled by the reduced matrix element $J_2$.  Second, a dimerized lattice can also be obtained in the setup \cite{Tarruell2012,Greif2013,Uehlinger2013}, where the dimerized honeycomb lattice sketched in \fig{fig:lattice} is created due to the interference of two collinear phase-shifted laser beams. Here, an additional shaking is not required but the control of the tunneling parameters is not independent. In the first case it is, e.g., possible to achieve negative tunneling matrix elements.

\begin{figure}[t]
\includegraphics[width=\linewidth]{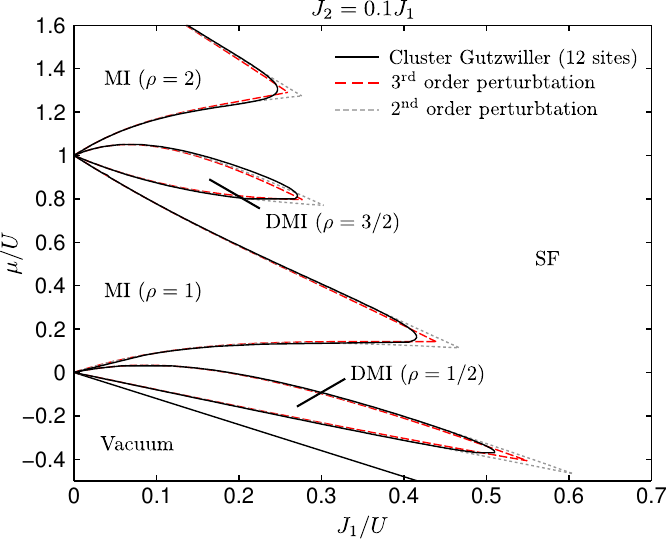} \centering
\caption{Phase diagram of the driven honeycomb lattice with $J_1 = 10  J_2$. For shallow lattices, i.e., large tunneling $J_1$ and $J_2$ the superfluid (SF) is favorable. For deep lattices the common Mott phases with integer fillings (MI) appear accompanied by dimerized Mott insulator phases with fractional fillings (DMI) in between.}
\label{fig:PD}
\end{figure}

The phase diagram for this setup is  shown in \fig{fig:PD} for $J_1=10 J_2$ as a function of the chemical potential $\mu$ and the tunneling energy $J_1$ in units of the repulsive on-site interaction $U$. Both the Mott insulator phases (MI) and the dimerized Mott insulator phases (DMI) are characterized by a vanishing superfluid order parameter $\psi=0$. In the remaining regions of the phase diagram with $\psi\neq0$ the particles are in a superfluid state. 
In the Mott insulator phase, the particles are localized at individual lattice sites as depicted in \fig{fig:lattice}b. In contrast, the dimerized Mott insulator phase with fractional filling $\rho=1/2$, $3/2$, $...$  is characterized by a delocalization of particles within the dimers (\fig{fig:lattice}c). In the limit of fully separated dimers ($J_2\to 0$), the local ground state on a dimer for filling $\rho=1/2$ reads $\ket{s}=\frac{1}{\sqrt{2}}(\ket{1,0}+\ket{0,1})$, where $\ket{n_\text{L},n_\text{R}}$ denotes the occupation of left and right dimer sites. 
In the dimerized insulator phase the local particle-number fluctuations $(\Delta n)^2$ are large, whereas they are strongly suppressed in a conventional Mott insulator, which can be used to distinguish between the phases.
Thus, the single-site resolution available in a quantum-gas microscope experiment would reveal a random distribution on individual lattice sites but a fixed integer occupation when summing up both dimer sites. When the tunneling matrix element $J_1$ is negative, which can be achieved with the lattice shaking technique, the antisymmetric state $\ket{a} = \frac{1}{\sqrt{2}} (\ket{1,0} - \ket{0,1})$ becomes the dimer ground state resembling a $p$-orbital. Hence, we can understand the setup as a square lattice of $p$-orbitals. In the superfluid phase, a negative tunneling matrix element leads to an alternating sign of the superfluid order parameter, which can be mapped onto a classical spin model and is therefore referred to as antiferromagnetic coupling. The phases align due to the positive dimer coupling $J_2 > 0$ such that the sign of the dimer wave function is the same along the respective bonds (\fig{fig:lattice}d), which minimizes the tunneling energy. This alignment leads to a stripe order of the superfluid order parameter. In the Mott state where the superfluid order parameter vanishes, this alignment persists in the nearest-neighbor correlations. A suitable gauge transformation maps the symmetric (ferromagnetic) onto the antisymmetric ground state which is not frustrated. Therefore, we restrict ourselves to the symmetric case for $J_1>0$ in the following. We should stress however that the aforementioned equivalence means that the dimerized lattice resembles $p$-orbital-like physics even without antiferromagnetic driving of the lattice.

\begin{figure*}[t]
\includegraphics[width=.8\linewidth]{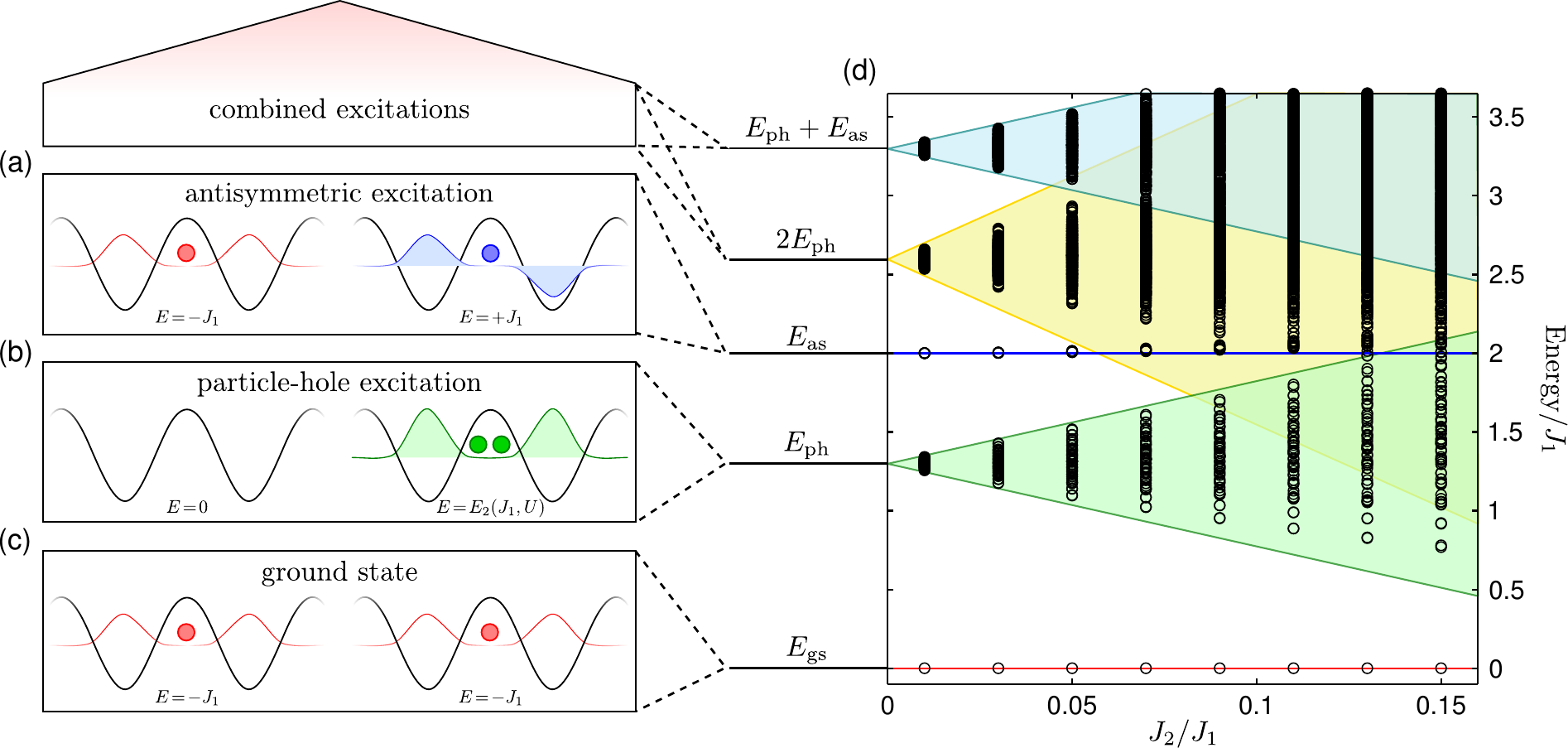} \centering
\caption{Fundamental excitations in the dimerized Mott insulator. (c) In the ground state the dimers are occupied by the symmetric state $\ket{s}$. (b) A particle-hole excitation with one empty and one doubly occupied dimer. (a) The fundamental excitation where one dimer is in an anti-symmetric state $\ket{a}$.
(d) Excitation spectrum of the dimerized Mott insulator at $J_1=0.2 U$. The shaded bands depict first-order strong-coupling perturbation results corresponding to the fundamental excitations depicted on the left. The black circles are numerical data using a cluster mean-field approach (16 sites). The discrepancy is caused by higher-order tunneling, the finite cluster size,  as well as the negligence of antisymmetric states in the perturbation approach. \label{fig:spectrum}}
\end{figure*}

The solid lines in the phase diagram in \fig{fig:PD} are computed using a cluster mean-field approach with a cluster size of 12 sites, as described in Sec.~\ref{sec:CMF}. Due to the structure of the dimerized state, conventional mean-field theory is not able to capture the dimerized Mott insulator phases. As a second approach, we apply  strong-coupling perturbative expansion to third order (dashed lines) which allows analytical results for the phase diagram. This approach is detailed below in Sec.~\ref{sec:PT}, where we also give the explicit expression for the second-order perturbation (dotted lines) as a reasonable approximation.
Both methods use the tight-binding Bose-Hubbard Hamiltonian
\begin{equation}\begin{split}\label{eq:H_full}
 \hat H =  \frac{U}{2}\sum_i \hat n_i (\hat n_i - 1)- J_1 \sum_{\langle i,j \rangle} \hat b_i^\dagger \hat b_{j} - J_2 \sum_{\langle\!\langle i,j \rangle\!\rangle} \hat b_i^\dagger \hat b_{j}. \end{split}\end{equation}
Here, the brackets $\langle i,j \rangle$ denote sites $i$ and $j$ on the same dimer connected via $J_1$, while sites $\langle\!\langle i,j \rangle\!\rangle$ are nearest-neighbors sites  connected via $J_2$ on different dimers. The repulsive on-site interaction is denoted by $U$ with the particle number operator $\hat n_i=\hat b_i^\dagger \hat b_i$. 

\section{Excitation spectrum} \label{sec:spectrum}
Due to the internal structure of the dimerized Mott insulator state, its fundamental excitations differ strongly from the normal Mott insulator phases. Therefore the excitation spectrum allows distinguishing between the two insulating phases. 
In the Mott insulator, the lowest excitation is the creation of a particle-hole pair resulting in an empty and a doubly occupied site. The corresponding excitation energy is the additional on-site interaction $U$. The excitation spectrum of the fractional insulator with $\rho=1/2$ is not $U$-gaped since each dimer offers an empty site, where the particle from a neighboring dimer can tunnel to. Even for dimerized Mott insulators with higher fillings the $U$-gap vanishes.

In \fig{fig:spectrum}c the ground-state configuration of the dimerized Mott insulator is depicted for two unit cells, each being populated by the symmetric state $\ket{s}$.
There are two different fundamental excitations both on the order of $2J_1$. First, the particle-hole excitation $E_\ph$, where one particle is excited by hopping to a neighboring dimer as depicted in \fig{fig:spectrum}b. The excitation energy corresponds to the loss of delocalization energy $J_1$ within the empty dimer reduced by the interaction energy on the doubly occupied dimer. Second, a particle can be excited within the same dimer from the symmetric ground state $\ket{s}$ to the antisymmetric state $\ket{a}$ associated with the energy $E_\text{as}=2J_1$ (see \fig{fig:spectrum}a).

\section{Harmonic confinement} \label{sec:inhomog}
\begin{figure*}[t]
\includegraphics[width=0.9\linewidth]{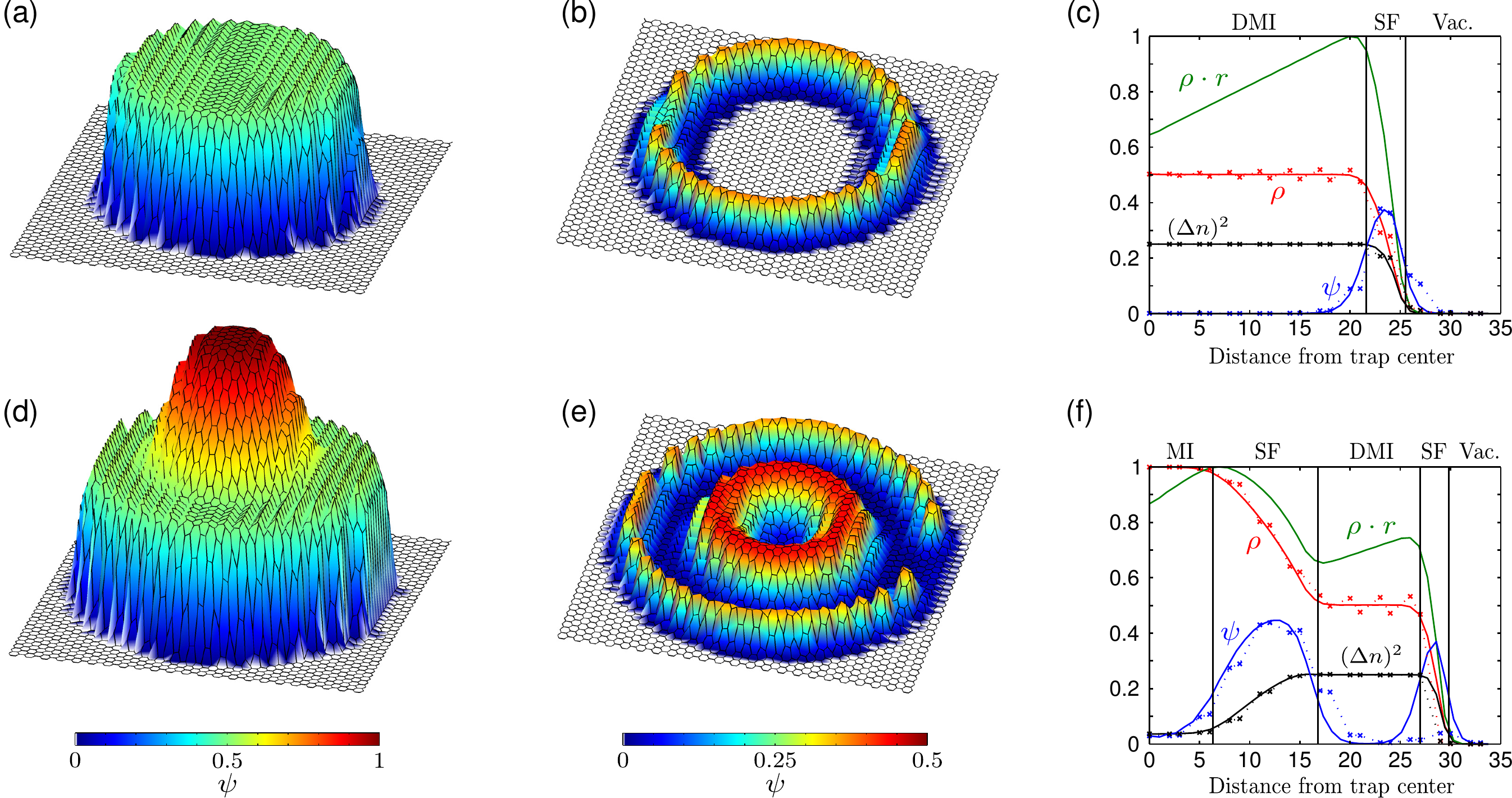}
\caption{Driven honeycomb lattice in a trap with a frequency of $f=32\, \mathrm{Hz}$ with $N=680$ atoms (a-c) and $f=35\, \mathrm{Hz}$ with $N=1160$ atoms (d-f). (a, d) The density of the atoms forms a wedding-cake structure with half-integer steps showing  the dimerized Mott insulator ($\rho=\frac12$) and the Mott insulator ($\rho=1$). (b, e) The order parameter $\psi = \braket{\hat b}$ characterizing the insulating phases with vanishing $\psi$ divided by superfluid rings. (c, f) Cuts through the center of the trap perpendicular (solid lines) and parallel (markers) to the axis of the dimers. The vanishing order parameter (blue lines) coincides with the density plateaus (red lines). The local particle-number fluctuations $(\Delta n)^2$ (black line) in connection with the gradient of the confining potential induces a density modulation in the dimerized Mott phase (red markers). The vertical black lines indicate the phase boundaries in the local density approximation.}
\label{fig:inhomog}
\end{figure*}

The excitation spectrum as a function of $J_2/J_1$ is depicted in \fig{fig:spectrum}d in units of the tunneling energy $J_1$ for $J_1=0.2U$. The black markers represent the numerical data calculated with the cluster mean-field approach using 16 sites within the insulating phase with $\psi=0$. The shaded areas indicate the results of first-order strong-coupling perturbation theory. The antisymmetric excitation $E_\text{as}$ is not broadened within first-order perturbation and is only affected by higher-order processes. In the limit $J_2 \rightarrow 0$, a particle-hole excitation for an insulating phase with $n$ particles per dimer has the energy  $E_\ph=E_{n+1}+E_{n-1} - 2E_n$. Thus, the particle-hole energy  for the fractional insulator with $n=1$ is $E_\ph = E_2-2 E_1\approx 2J_1-4J_1^2/U$. For finite $J_2$, first order perturbation (blue shaded area) leads to a delocalization of the particle and the hole. The minimum and maximum of the emerging band are given by
\begin{equation}
	E_\ph = E_2-2 E_1 \pm 4 J_2 (j_1^2+j_{2}^2). 
\end{equation}
where $j_n^2+j_{n+1}^2\approx 1$ for $U\gg J_1$ (see Sec.~\ref{sec:PT} for details).

The excitation gap $E_\mathrm{ph} \approx J_1$ gives an estimate of the required temperature for the dimerized phase $T \approx E_\mathrm{ph}/k_\mathrm{B}$, where $k_\mathrm{B}$ is the Boltzmann constant. Assuming the experimental parameters given in section \ref{sec:inhomog} and $V=6.5\, E_\mathrm{rec}$ we obtain $T \approx 20\, \mathrm{nK}$. In general, this value increases with $J_1$ and decreases with the ratio $J_2/J_1$. Alternatively, a larger value of the scattering length $a_s/a$, where $a$ is the lattice constant, allows for a larger value of $J_1$, increasing the excitation gap.

The higher excitations in the energy spectrum are combinations of the two fundamental excitations described above, i.e. two particle-hole excitations $2E_\ph$, two asymmetric excitations $2E_\text{as}$, and a combination of both $E_\ph+E_\text{as}$.
The numerical spectrum is distorted by higher-order tunneling processes and the interaction between individual excitations.
Due to the finite size of the cluster the band width is reduced, which is in particular noticeable for two particle-hole excitations due to the  limited possibilities to delocalize. The $U$-excitation at $E\approx 5J_1$ corresponding to a doubly occupied site lies in the continuous part of the spectrum for most parameters. 

In an experiment, the lattice modulation technique proposed in \quote{Konabe2006} can be applied for the direct observation of the dimerized Mott insulator phase by addressing these fundamental excitations. The excitation gap on the order of $2 J_1$ is therefore the characteristic signature of the fractional insulator and  could serve as prove for its experimental realization.

In this section, the question is addressed whether the dimerized Mott insulator in the honeycomb lattice can be observed in an experimental setup, where the optical lattice is superimposed by a harmonic confinement, leading to a spatially varying chemical potential $\mu(\mathbf{r})$. The relatively small extend of the dimerized Mott insulator phase in the phase diagram might suggest that a dominating part of the atoms are in the superfluid or in the Mott insulator with $\rho=1$ coexisting with the dimerized phase. 
The cluster mean-field approach (see Sec.~\ref{sec:CMF}) allows the simulation of a two-dimensional lattice of realistic size by iteratively moving the cluster through the lattice \cite{Pisarski2011,Luhmann2013}. This introduces a site-dependent mean-field, where at every iteration the \textit{local} order parameter is updated until the results converge. If the ratio $J_1/U$ is considerably smaller than the tip position of the dimerized insulator phase, the results are influenced only to a minor degree by the cluster size. Using six-site clusters, we can determine the extent of the phases in a lattice with harmonic confinement accurately. In \fig{fig:inhomog} the results are shown for $J_1=0.1 U$, $J_2=0.1 J_1$ and a chemical potential $\mu_c$ in the center of the trap. As an example, for $^{87}\mathrm{Rb}$ with a scattering length of $a_s \approx 100\, a_0$, this corresponds to a lattice depth of $V=9.5\, E_\mathrm{rec}$ \cite{SoltanPanahi2011,Luhmann2014}, where $a_0$ is the Bohr radius, $E_\mathrm{rec}=\frac{\hbar^2 k^2}{2m}$ is the recoil energy, $k=\frac{2\pi}{\lambda}$ is the wave vector of the lattice beams with a wavelength of $\lambda=830\, \mathrm{nm}$ and $m$ is the atomic mass of $^{87}\mathrm{Rb}$. For each lattice site we apply the local density-approximation  $\mu(\mathbf{r}_i)=\mu_\text{c}-V_\text{trap}(\mathbf{r}_i)$ with a harmonic trapping potential $V_\text{trap}$ and trap frequencies of $32\, \mathrm{Hz}$ (a-c) to $35\, \mathrm{Hz}$ (d-f).

For $\mu_\text{c}=0.02 U$, only the dimerized insulator persists (\fig{fig:inhomog}a-c), whereas for $\mu_\text{c}=0.1U$ we observe the coexistence of normal and dimerized Mott insulator (\fig{fig:inhomog}d-f). The superfluid order parameter $\psi$ shown in \fig{fig:inhomog}b,e vanishes in the insulating phases and increases to a finite value in between forming superfluid rings. For a threshold of $\Psi<0.1$, we find that the dimerized Mott insulator phase is occupied by 490 of a total of 680 atoms (a-c) and 400 of 1160 atoms (d-f), respectively. The density in \fig{fig:inhomog}a,d shows the typical wedding cake structure but with half-integer steps. In the dimerized Mott insulator phase with $\rho=1/2$, a periodic density modulation along the dimer axis appears. While the average density on a dimer is fixed to $\rho=1/2$, the density on the two dimer sites adjust itself according to the gradient of the chemical potential along the dimers. The persistence of the insulating phase despite this strong impact of the gradient illustrates its robustness.

 A cut through the trap along the dimer axes is shown in \fig{fig:inhomog}c, f. The density profile (red lines) and the superfluid order parameter (blue lines) clearly indicate the Mott insulator and dimerized Mott insulator regime. This agrees well with the expectation from the phase diagram in \fig{fig:PD} for infinite size (for $\mu<\mu_c$). The dimerized Mott insulator plateaus in the density profile show an oscillating behavior along the dimer axis which is caused by the trap as discussed above. The green lines represent the total particle number and indicate a comparatively large occupation of the dimerized phase in the case of \fig{fig:inhomog}f. 

The local particle-number fluctuations $(\Delta n)^2$ are shown as black lines and demonstrate the expected large value of $(\Delta n)^2 = 0.25$ in the dimerized insulator phase. The large particle-number fluctuations in combination with the vanishing order parameter characterize the dimerized Mott insulator phase, whereas in the conventional Mott insulator all fluctuations are suppressed.

\section{Perturbation theory} \label{sec:PT}
The phase diagram \fig{fig:PD} as well as the excitation spectrum \fig{fig:spectrum}d can be approximated with the strong-coupling perturbative expansion technique \cite{Freericks1994,Freericks1996,Buonsante2004b,Buonsante2004c,Buonsante2005a}. For this we recast the full Hamiltonian \eqref{eq:H_full} to an effective model with dimer unit cells. As a first step we find the eigenstates $|n\rangle$ of $n$ particles in a dimer with energies $E_n$. In contrast to \quote{Buonsante2005a}, we can restrict the calculations to the respective symmetric ground states, due to the large ratio $J_1/J_2$. This simplifies the approach significantly and allows us to include perturbations up to third order. The energies for the lowest values of $n$ read $E_0=0$, $E_1=-J_1$, $E_2=\frac{1}{2}(U-\sqrt{U^2+16 J_1^2})$. This approximation is valid as long as the perturbation $z J_2$ with the coordination number $z=4$ is much smaller than the energy $E_n^{(1)}- E_n^{(0)}$, i.e.,  
$z J_2\ll U, J_1 $.
The coupling between neighboring dimers is given by the operator
\begin{equation}
 \hat J_2 = - J_2 \sum_{\langle i,j \rangle} \hat d_i^\dagger \hat d_{j},
\end{equation}
where the brackets $\langle i,j \rangle$ label neighboring dimers $i, j$ instead of sites and the operators $\hat d_i (\hat d_i^\dagger)$ annihilate (create) a particle on a dimer. More precisely, for the annihilation of a particle on a dimer the operator is given by the projection on the $n$-particle ground states
\begin{equation}
 \hat d = \sum_n |n-1\rangle\langle n-1 | \hat b |n\rangle\langle n |,
\end{equation}
where $\hat b$ acts on one of the two equivalent sites of the dimer. As a compact notation we define the coupling parameter $j_n = \langle n-1 | \hat b |n\rangle$. For a conventional single-site lattice model it is $j_n=\sqrt{n}$, whereas here $j_n$ depends on the exact form of the ground states that are functions of $J_1/U$.
With the above restriction we obtain an effective lattice model 
\begin{equation}\label{eq:H_eff}
 \hat H_\mathrm{eff} = \sum_i (E_{\hat n_i} - \mu \hat n_i) - J_2 \sum_{\langle i,j \rangle} \hat d_i^\dagger \hat d_{j}.
\end{equation}
Within this dimerized model, all insulating phases are treated on the same footing as product states of dimer ground states $|n\rangle$. Odd fillings $n$ correspond to dimerized insulators and even $n$ to conventional Mott insulators.

We now discuss the first lowest three orders of perturbation by inter-dimer tunneling $J_2$. In the unperturbed case of $J_2=0$, the creation of a particle (hole) excitation is associated with the energy
\begin{align}
E_\mathrm{p}^{(0)} &= E_{n + 1} - E_n - \mu \\
E_\mathrm{h}^{(0)} &= E_{n - 1} - E_n + \mu,
\end{align}
with the chemical potential $\mu$. In this case, only insulating phases exist and their phase boundaries $\mu_{p|h}^{(0)}$ can be obtained from the condition $E_{p|h}=0$, where the chemical potential compensates for the energy of one additional or missing particle per dimer.

In first order perturbation, the delocalization of a  particle or a hole over the lattice results in lower and upper energy bounds
\begin{align}
 E_{\mathrm{p}\pm}^{(1)} &= E_\mathrm{p}^{(0)} \pm z J_2 j_{n+1}^2 \\
 E_{\mathrm{h}\pm}^{(1)} &= E_\mathrm{h}^{(0)} \pm z J_2 j_n^2.
\end{align}
Thus, the energy band for a particle-hole excitations lies between $E_{\ph\pm} = E_{\mathrm{p}\pm}^{(1)}+ E_{\mathrm{h}\pm}^{(1)}$ for one and $2 E_{\ph\pm}$ for two excitations indicated in figure \fig{fig:spectrum}d (colored areas). In addition, the asymmetric excitation that goes beyond the effective model \eqref{eq:H_eff} is indicated as  unbroadened line.

The coupling between the dimers gives rise to superfluid phases surrounding the insulating phases. The phase boundaries shown in \fig{fig:PD} are obtained from the minimum energy of the excitations.  In first order perturbation they read 
\begin{align}
\mu_\mathrm{p}^{(1)} &= E_{n+1} - E_n - z J_2 j_{n+1}^2 \\
\mu_\mathrm{h}^{(1)} &= E_n - E_{n-1} + z J_2 j_n^2.
\end{align}

For the determination of second- and third-order energy, one has to include  processes with amplitudes on the order of $J_2^2$ and $J_2^3$. The phase boundaries in second-order perturbation read
\begin{align}
\begin{split} \label{eq_mup2}
 \mu_\mathrm{p}^{(2)} = \mu_\mathrm{p}^{(1)} &+ 8 \frac{(J_2 j_n j_{n+1})^2}{E_{n+1}+E_{n-1}-2E_n} \\
&- 4 \frac{(J_2 j_n j_{n+2})^2}{E_{n+2}+E_{n-1}-E_n-E_{n+1}}
\end{split} \\
\begin{split}
\mu_\mathrm{h}^{(2)} = \mu_\mathrm{h}^{(1)} &+ 4 \frac{(J_2 j_n j_{n+1})^2}{E_{n+1}+E_{n-1}-2E_n} \\
&+ 4 \frac{(J_2 j_{n-1} j_{n+1})^2}{E_{n-2}+E_{n+1}-E_n-E_{n-1}}
\end{split}.
\end{align}
The first additional term in \eqref{eq_mup2} accounts for the second-order energy of the insulator that is obtained by processes via a virtual particle-hole excitations. When a particle excitation is present, these bidirectional processes are inhibited on neighboring bonds leading to a factor of $2 z = 8$. They are partly substituted by processes via the intermediate state where a particle tunnels onto the excitation, which is captured by the second additional term. Analogously, the energy of a hole-excitation relative to the insulator accounts for all possible second-order processes.

We compute the phase diagram up to third order and find a good agreement with the cluster mean-field approach detailed below. The pointy tip of the insulator lobes is due to the vanishing energy of particle-hole excitations at the crossing point of the phase boundaries. Here, the perturbation series cannot be limited to finite-order processes.

\section{Cluster Gutzwiller theory} \label{sec:CMF}
Our numerical simulations are performed using a cluster mean-field approach \cite{Buonsante2004b,Jain2004,Hen2009,Pisarski2011,McIntosh2012,Yamamoto2012b,Luhmann2013}. The cluster approach allows capturing phases with strong short-range correlations such as fractional insulators formed on unit cells covering more than one site. Conventional single-site mean-field approaches such as the Gutzwiller approach are not capable of finding the dimerized insulator phases. Furthermore the cluster mean-field approach is well suited to obtain results for inhomogeneous systems as discussed in \sec{sec:inhomog} and gives detailed information on the local excitation spectrum presented in \sec{sec:spectrum}. 

The cluster Gutzwiller approach decouples a cluster of sites from the rest of the lattice and couples it to a mean-field at its borders. The latter is determined from the exact diagonalization of the cluster and is updated in an iterative process. This allows taking into account local correlations exactly and thereby gives far more precise results than conventional mean-field methods. The improvement is especially pronounced for lattices with a small number of nearest neighbors, such as hexagonal lattices. In the many-particle cluster basis $\ket{N}$ the Hamiltonian matrix 
\begin{equation}
	\hat H_{MN}=\bra{M} \hat H_\text{cluster}   + \hat H_\text{boundary}  \ket{N} 
	\label{eq:HamiltionanMatrix}
\end{equation}
decomposes in two parts describing the cluster according to Eq.~\eqref{eq:H_full} and its boundary. The Hamiltonian $\hat H_\text{boundary} $ describes the coupling of sites at the boundary $\sigma$ of the cluster to sites outside the cluster and reads
\begin{equation}
 \hat H_\mathrm{boundary}=-J_2 \sum_{i \in \sigma} \nu_i \hat b^\dagger_i \braket{\hat b} + c.c.,
\end{equation}
where $\nu_i$ is the number of mean-field bonds at site $i$. The superfluid order parameter $\psi = \braket{\hat b}$ at the boundary is obtained from the innermost site in the cluster. 
When at least one of the two tunneling matrix elements is negative, the order parameter $\psi$ shows an alternating sign as depicted in \fig{fig:lattice}d. 
Note that we apply periodic boundary conditions along one direction of the clusters, which increases the ratio of inner cluster bonds to mean-field bonds. We use clusters of up to 16 sites and restrict the basis $\ket{N}$ further using cut-offs on the fluctuations and the number of particles per site (for further details see \quote{Luhmann2013}). We carefully checked for convergence.

The excitation spectrum obtained from the cluster mean-field approach agrees well with the perturbation theory. The reduced band width of the band of two particle-hole excitations is due to the finite size of the cluster, as the two excitations already spread over four dimers on a eight-dimer cluster, which limits the possibilities for delocalization. 

The phase diagram matches the predictions of the perturbation theory well for all phases ($0 \leq n \leq 4$). At the tips of the lobes, the perturbation theory is not expected to give precise results due to the vanishing energy of particle-hole excitations. We expect further deviations due to the restriction to the symmetric ground states on each dimer in the perturbation theory approach. The cluster mean-field approach on the other hand is restricted to a finite cluster size. The good agreement between the two methods indicates that the approximations are justified and the true phase boundary is well approximated. 

\section{Conclusions}

In conclusion, we have shown that when anisotropic tunneling is introduced to optical honeycomb lattices, dimerized Mott insulator phases with fractional fillings appear.
In these phases, the superfluid order parameter vanishes but large particle number fluctuations persist on the individual lattice sites. 
 We have calculated the phase diagram using two different approaches, namely a cluster Gutzwiller approach and the strong-coupling perturbation expansion technique, and found excellent agreement. The former method allows us to study the excitation spectrum, which allows the distinction of normal and dimerized Mott insulator. In a possible experiment with a harmonic confinement the dimerized insulator is formed by a large fraction of the atoms and should therefore be observable. Therefore, optical honeycomb lattice experiments should be well suited to realize and probe the proposed dimerized Mott insulator phase, which to our best knowledge has not been measured experimentally so far.
When driving an optical honeycomb lattice the intra-dimer tunneling coupling can be tuned negative to realize $p$-orbital like physics.

After submission, we became aware of \quote{Gawryluk2013} discussing a similar setup.

\section{Acknowledgments}
We thank K. Sengstock, J.Struck and M. Weinberg for helpful discussions. We acknowledge funding by the Deutsche Forschungsgemeinschaft (grants SFB 925 and GRK 1355).


\begin{thebibliography}{48}
\expandafter\ifx\csname natexlab\endcsname\relax\def\natexlab#1{#1}\fi
\expandafter\ifx\csname bibnamefont\endcsname\relax
  \def\bibnamefont#1{#1}\fi
\expandafter\ifx\csname bibfnamefont\endcsname\relax
  \def\bibfnamefont#1{#1}\fi
\expandafter\ifx\csname citenamefont\endcsname\relax
  \def\citenamefont#1{#1}\fi
\expandafter\ifx\csname url\endcsname\relax
  \def\url#1{\texttt{#1}}\fi
\expandafter\ifx\csname urlprefix\endcsname\relax\def\urlprefix{URL }\fi
\providecommand{\bibinfo}[2]{#2}
\providecommand{\eprint}[2][]{\url{#2}}

\bibitem[{\citenamefont{Jaksch et~al.}(1998)\citenamefont{Jaksch, Bruder,
  Cirac, Gardiner, and Zoller}}]{Jaksch1998}
\bibinfo{author}{\bibfnamefont{D.}~\bibnamefont{Jaksch}},
  \bibinfo{author}{\bibfnamefont{C.}~\bibnamefont{Bruder}},
  \bibinfo{author}{\bibfnamefont{J.~I.} \bibnamefont{Cirac}},
  \bibinfo{author}{\bibfnamefont{C.~W.} \bibnamefont{Gardiner}},
  \bibnamefont{and} \bibinfo{author}{\bibfnamefont{P.}~\bibnamefont{Zoller}},
  \bibinfo{journal}{Phys. Rev. Lett.} \textbf{\bibinfo{volume}{81}},
  \bibinfo{pages}{3108} (\bibinfo{year}{1998}).

\bibitem[{\citenamefont{{Greiner} et~al.}(2002)\citenamefont{{Greiner},
  {Mandel}, {Esslinger}, {H{\"a}nsch}, and {Bloch}}}]{Greiner2002}
\bibinfo{author}{\bibfnamefont{M.}~\bibnamefont{{Greiner}}},
  \bibinfo{author}{\bibfnamefont{O.}~\bibnamefont{{Mandel}}},
  \bibinfo{author}{\bibfnamefont{T.}~\bibnamefont{{Esslinger}}},
  \bibinfo{author}{\bibfnamefont{T.~W.} \bibnamefont{{H{\"a}nsch}}},
  \bibnamefont{and} \bibinfo{author}{\bibfnamefont{I.}~\bibnamefont{{Bloch}}},
  \bibinfo{journal}{\nat} \textbf{\bibinfo{volume}{415}}, \bibinfo{pages}{39}
  (\bibinfo{year}{2002}).

\bibitem[{\citenamefont{Guidoni et~al.}(1997)\citenamefont{Guidoni, Trich\'e,
  Verkerk, and Grynberg}}]{Guidoni1997}
\bibinfo{author}{\bibfnamefont{L.}~\bibnamefont{Guidoni}},
  \bibinfo{author}{\bibfnamefont{C.}~\bibnamefont{Trich\'e}},
  \bibinfo{author}{\bibfnamefont{P.}~\bibnamefont{Verkerk}}, \bibnamefont{and}
  \bibinfo{author}{\bibfnamefont{G.}~\bibnamefont{Grynberg}},
  \bibinfo{journal}{Phys. Rev. Lett.} \textbf{\bibinfo{volume}{79}},
  \bibinfo{pages}{3363} (\bibinfo{year}{1997}).

\bibitem[{\citenamefont{Peil et~al.}(2003)\citenamefont{Peil, Porto, Tolra,
  Obrecht, King, Subbotin, Rolston, and Phillips}}]{Peil2003}
\bibinfo{author}{\bibfnamefont{S.}~\bibnamefont{Peil}},
  \bibinfo{author}{\bibfnamefont{J.~V.} \bibnamefont{Porto}},
  \bibinfo{author}{\bibfnamefont{B.~L.} \bibnamefont{Tolra}},
  \bibinfo{author}{\bibfnamefont{J.~M.} \bibnamefont{Obrecht}},
  \bibinfo{author}{\bibfnamefont{B.~E.} \bibnamefont{King}},
  \bibinfo{author}{\bibfnamefont{M.}~\bibnamefont{Subbotin}},
  \bibinfo{author}{\bibfnamefont{S.~L.} \bibnamefont{Rolston}},
  \bibnamefont{and} \bibinfo{author}{\bibfnamefont{W.~D.}
  \bibnamefont{Phillips}}, \bibinfo{journal}{Phys. Rev. A}
  \textbf{\bibinfo{volume}{67}}, \bibinfo{pages}{051603}
  (\bibinfo{year}{2003}).

\bibitem[{\citenamefont{Santos et~al.}(2004)\citenamefont{Santos, Baranov,
  Cirac, Everts, Fehrmann, and Lewenstein}}]{Santos2004}
\bibinfo{author}{\bibfnamefont{L.}~\bibnamefont{Santos}},
  \bibinfo{author}{\bibfnamefont{M.~A.} \bibnamefont{Baranov}},
  \bibinfo{author}{\bibfnamefont{J.~I.} \bibnamefont{Cirac}},
  \bibinfo{author}{\bibfnamefont{H.-U.} \bibnamefont{Everts}},
  \bibinfo{author}{\bibfnamefont{H.}~\bibnamefont{Fehrmann}}, \bibnamefont{and}
  \bibinfo{author}{\bibfnamefont{M.}~\bibnamefont{Lewenstein}},
  \bibinfo{journal}{Phys. Rev. Lett.} \textbf{\bibinfo{volume}{93}},
  \bibinfo{pages}{030601} (\bibinfo{year}{2004}).

\bibitem[{\citenamefont{Anderlini et~al.}(2007)\citenamefont{Anderlini, Lee,
  Brown, Sebby-Strabley, Phillips, and Porto}}]{Anderlini2007}
\bibinfo{author}{\bibfnamefont{M.}~\bibnamefont{Anderlini}},
  \bibinfo{author}{\bibfnamefont{P.~J.} \bibnamefont{Lee}},
  \bibinfo{author}{\bibfnamefont{B.~L.} \bibnamefont{Brown}},
  \bibinfo{author}{\bibfnamefont{J.}~\bibnamefont{Sebby-Strabley}},
  \bibinfo{author}{\bibfnamefont{W.~D.} \bibnamefont{Phillips}},
  \bibnamefont{and} \bibinfo{author}{\bibfnamefont{J.~V.} \bibnamefont{Porto}},
  \bibinfo{journal}{Nature} \textbf{\bibinfo{volume}{448}},
  \bibinfo{pages}{452} (\bibinfo{year}{2007}).

\bibitem[{\citenamefont{Trotzky et~al.}(2008)\citenamefont{Trotzky, Cheinet,
  F\"olling, Feld, Schnorrberger, Rey, Polkovnikov, Demler, Lukin, and
  Bloch}}]{Trotzky2008}
\bibinfo{author}{\bibfnamefont{S.}~\bibnamefont{Trotzky}},
  \bibinfo{author}{\bibfnamefont{P.}~\bibnamefont{Cheinet}},
  \bibinfo{author}{\bibfnamefont{S.}~\bibnamefont{F\"olling}},
  \bibinfo{author}{\bibfnamefont{M.}~\bibnamefont{Feld}},
  \bibinfo{author}{\bibfnamefont{U.}~\bibnamefont{Schnorrberger}},
  \bibinfo{author}{\bibfnamefont{A.~M.} \bibnamefont{Rey}},
  \bibinfo{author}{\bibfnamefont{A.}~\bibnamefont{Polkovnikov}},
  \bibinfo{author}{\bibfnamefont{E.~A.} \bibnamefont{Demler}},
  \bibinfo{author}{\bibfnamefont{M.~D.} \bibnamefont{Lukin}}, \bibnamefont{and}
  \bibinfo{author}{\bibfnamefont{I.}~\bibnamefont{Bloch}},
  \bibinfo{journal}{Science} \textbf{\bibinfo{volume}{319}},
  \bibinfo{pages}{295} (\bibinfo{year}{2008}).

\bibitem[{\citenamefont{Sebby-Strabley
  et~al.}(2006)\citenamefont{Sebby-Strabley, Anderlini, Jessen, and
  Porto}}]{Sebby-Strabley2006}
\bibinfo{author}{\bibfnamefont{J.}~\bibnamefont{Sebby-Strabley}},
  \bibinfo{author}{\bibfnamefont{M.}~\bibnamefont{Anderlini}},
  \bibinfo{author}{\bibfnamefont{P.~S.} \bibnamefont{Jessen}},
  \bibnamefont{and} \bibinfo{author}{\bibfnamefont{J.~V.} \bibnamefont{Porto}},
  \bibinfo{journal}{Phys. Rev. A} \textbf{\bibinfo{volume}{73}},
  \bibinfo{pages}{033605} (\bibinfo{year}{2006}).

\bibitem[{\citenamefont{Wirth et~al.}({2011})\citenamefont{Wirth,
  \"Olschl\"ager, and Hemmerich}}]{Wirth2011}
\bibinfo{author}{\bibfnamefont{G.}~\bibnamefont{Wirth}},
  \bibinfo{author}{\bibfnamefont{M.}~\bibnamefont{\"Olschl\"ager}},
  \bibnamefont{and}
  \bibinfo{author}{\bibfnamefont{A.}~\bibnamefont{Hemmerich}},
  \bibinfo{journal}{{Nature Phys.}} \textbf{\bibinfo{volume}{{7}}},
  \bibinfo{pages}{{147}} (\bibinfo{year}{{2011}}).

\bibitem[{\citenamefont{Jo et~al.}(2012)\citenamefont{Jo, Guzman, Thomas,
  Hosur, Vishwanath, and Stamper-Kurn}}]{Jo2012}
\bibinfo{author}{\bibfnamefont{G.-B.} \bibnamefont{Jo}},
  \bibinfo{author}{\bibfnamefont{J.}~\bibnamefont{Guzman}},
  \bibinfo{author}{\bibfnamefont{C.~K.} \bibnamefont{Thomas}},
  \bibinfo{author}{\bibfnamefont{P.}~\bibnamefont{Hosur}},
  \bibinfo{author}{\bibfnamefont{A.}~\bibnamefont{Vishwanath}},
  \bibnamefont{and} \bibinfo{author}{\bibfnamefont{D.~M.}
  \bibnamefont{Stamper-Kurn}}, \bibinfo{journal}{Phys. Rev. Lett.}
  \textbf{\bibinfo{volume}{108}}, \bibinfo{pages}{045305}
  (\bibinfo{year}{2012}).

\bibitem[{\citenamefont{Becker et~al.}(2010)\citenamefont{Becker,
  Soltan-Panahi, Kronj\"ager, D\"orscher, Bongs, and Sengstock}}]{Becker2010}
\bibinfo{author}{\bibfnamefont{C.}~\bibnamefont{Becker}},
  \bibinfo{author}{\bibfnamefont{P.}~\bibnamefont{Soltan-Panahi}},
  \bibinfo{author}{\bibfnamefont{J.}~\bibnamefont{Kronj\"ager}},
  \bibinfo{author}{\bibfnamefont{S.}~\bibnamefont{D\"orscher}},
  \bibinfo{author}{\bibfnamefont{K.}~\bibnamefont{Bongs}}, \bibnamefont{and}
  \bibinfo{author}{\bibfnamefont{K.}~\bibnamefont{Sengstock}},
  \bibinfo{journal}{New J. Phys.} \textbf{\bibinfo{volume}{12}},
  \bibinfo{pages}{065025} (\bibinfo{year}{2010}).

\bibitem[{\citenamefont{Soltan-Panahi et~al.}(2011)\citenamefont{Soltan-Panahi,
  Struck, Hauke, Bick, Plenkers, Meineke, Becker, Windpassinger, Lewenstein,
  and Sengstock}}]{SoltanPanahi2011}
\bibinfo{author}{\bibfnamefont{P.}~\bibnamefont{Soltan-Panahi}},
  \bibinfo{author}{\bibfnamefont{J.}~\bibnamefont{Struck}},
  \bibinfo{author}{\bibfnamefont{P.}~\bibnamefont{Hauke}},
  \bibinfo{author}{\bibfnamefont{A.}~\bibnamefont{Bick}},
  \bibinfo{author}{\bibfnamefont{W.}~\bibnamefont{Plenkers}},
  \bibinfo{author}{\bibfnamefont{G.}~\bibnamefont{Meineke}},
  \bibinfo{author}{\bibfnamefont{C.}~\bibnamefont{Becker}},
  \bibinfo{author}{\bibfnamefont{P.}~\bibnamefont{Windpassinger}},
  \bibinfo{author}{\bibfnamefont{M.}~\bibnamefont{Lewenstein}},
  \bibnamefont{and}
  \bibinfo{author}{\bibfnamefont{K.}~\bibnamefont{Sengstock}},
  \bibinfo{journal}{Nature Phys.} \textbf{\bibinfo{volume}{7}},
  \bibinfo{pages}{434} (\bibinfo{year}{2011}).

\bibitem[{\citenamefont{Soltan-Panahi
  et~al.}({2012})\citenamefont{Soltan-Panahi, L\"uhmann, Struck, Windpassinger,
  and Sengstock}}]{SoltanPanahi2012}
\bibinfo{author}{\bibfnamefont{P.}~\bibnamefont{Soltan-Panahi}},
  \bibinfo{author}{\bibfnamefont{D.-S.} \bibnamefont{L\"uhmann}},
  \bibinfo{author}{\bibfnamefont{J.}~\bibnamefont{Struck}},
  \bibinfo{author}{\bibfnamefont{P.}~\bibnamefont{Windpassinger}},
  \bibnamefont{and}
  \bibinfo{author}{\bibfnamefont{K.}~\bibnamefont{Sengstock}},
  \bibinfo{journal}{{Nature Phys.}} \textbf{\bibinfo{volume}{{8}}},
  \bibinfo{pages}{{71}} (\bibinfo{year}{{2012}}).

\bibitem[{\citenamefont{Tarruell et~al.}({2012})\citenamefont{Tarruell, Greif,
  Uehlinger, Jotzu, and Esslinger}}]{Tarruell2012}
\bibinfo{author}{\bibfnamefont{L.}~\bibnamefont{Tarruell}},
  \bibinfo{author}{\bibfnamefont{D.}~\bibnamefont{Greif}},
  \bibinfo{author}{\bibfnamefont{T.}~\bibnamefont{Uehlinger}},
  \bibinfo{author}{\bibfnamefont{G.}~\bibnamefont{Jotzu}}, \bibnamefont{and}
  \bibinfo{author}{\bibfnamefont{T.}~\bibnamefont{Esslinger}},
  \bibinfo{journal}{{\nat}} \textbf{\bibinfo{volume}{{483}}},
  \bibinfo{pages}{{302}} (\bibinfo{year}{{2012}}).

\bibitem[{\citenamefont{Greif et~al.}(2013)\citenamefont{Greif, Uehlinger,
  Jotzu, Tarruell, and Esslinger}}]{Greif2013}
\bibinfo{author}{\bibfnamefont{D.}~\bibnamefont{Greif}},
  \bibinfo{author}{\bibfnamefont{T.}~\bibnamefont{Uehlinger}},
  \bibinfo{author}{\bibfnamefont{G.}~\bibnamefont{Jotzu}},
  \bibinfo{author}{\bibfnamefont{L.}~\bibnamefont{Tarruell}}, \bibnamefont{and}
  \bibinfo{author}{\bibfnamefont{T.}~\bibnamefont{Esslinger}},
  \bibinfo{journal}{Science} \textbf{\bibinfo{volume}{340}},
  \bibinfo{pages}{1307} (\bibinfo{year}{2013}).

\bibitem[{\citenamefont{Uehlinger et~al.}(2013)\citenamefont{Uehlinger, Jotzu,
  Messer, Greif, Hofstetter, Bissbort, and Esslinger}}]{Uehlinger2013}
\bibinfo{author}{\bibfnamefont{T.}~\bibnamefont{Uehlinger}},
  \bibinfo{author}{\bibfnamefont{G.}~\bibnamefont{Jotzu}},
  \bibinfo{author}{\bibfnamefont{M.}~\bibnamefont{Messer}},
  \bibinfo{author}{\bibfnamefont{D.}~\bibnamefont{Greif}},
  \bibinfo{author}{\bibfnamefont{W.}~\bibnamefont{Hofstetter}},
  \bibinfo{author}{\bibfnamefont{U.}~\bibnamefont{Bissbort}}, \bibnamefont{and}
  \bibinfo{author}{\bibfnamefont{T.}~\bibnamefont{Esslinger}},
  \bibinfo{journal}{Phys. Rev. Lett.} \textbf{\bibinfo{volume}{111}},
  \bibinfo{pages}{185307} (\bibinfo{year}{2013}).

\bibitem[{\citenamefont{{L{\"u}hmann} et~al.}(2014)\citenamefont{{L{\"u}hmann},
  {J{\"u}rgensen}, {Weinberg}, {Simonet}, {Soltan-Panahi}, and
  {Sengstock}}}]{Luhmann2014}
\bibinfo{author}{\bibfnamefont{D.-S.} \bibnamefont{{L{\"u}hmann}}},
  \bibinfo{author}{\bibfnamefont{O.}~\bibnamefont{{J{\"u}rgensen}}},
  \bibinfo{author}{\bibfnamefont{M.}~\bibnamefont{{Weinberg}}},
  \bibinfo{author}{\bibfnamefont{J.}~\bibnamefont{{Simonet}}},
  \bibinfo{author}{\bibfnamefont{P.}~\bibnamefont{{Soltan-Panahi}}},
  \bibnamefont{and}
  \bibinfo{author}{\bibfnamefont{K.}~\bibnamefont{{Sengstock}}},
  \bibinfo{journal}{arXiv:1401.5961}  (\bibinfo{year}{2014}).

\bibitem[{\citenamefont{Zhu et~al.}(2007)\citenamefont{Zhu, Wang, and
  Duan}}]{Zhu2007}
\bibinfo{author}{\bibfnamefont{S.-L.} \bibnamefont{Zhu}},
  \bibinfo{author}{\bibfnamefont{B.}~\bibnamefont{Wang}}, \bibnamefont{and}
  \bibinfo{author}{\bibfnamefont{L.-M.} \bibnamefont{Duan}},
  \bibinfo{journal}{Phys. Rev. Lett.} \textbf{\bibinfo{volume}{98}},
  \bibinfo{pages}{260402} (\bibinfo{year}{2007}).

\bibitem[{\citenamefont{Wu and Das~Sarma}(2008)}]{Wu2008}
\bibinfo{author}{\bibfnamefont{C.}~\bibnamefont{Wu}} \bibnamefont{and}
  \bibinfo{author}{\bibfnamefont{S.}~\bibnamefont{Das~Sarma}},
  \bibinfo{journal}{Phys. Rev. B} \textbf{\bibinfo{volume}{77}},
  \bibinfo{pages}{235107} (\bibinfo{year}{2008}).

\bibitem[{\citenamefont{Chen and Wu}(2011)}]{Chen2011}
\bibinfo{author}{\bibfnamefont{Z.}~\bibnamefont{Chen}} \bibnamefont{and}
  \bibinfo{author}{\bibfnamefont{B.}~\bibnamefont{Wu}}, \bibinfo{journal}{Phys.
  Rev. Lett.} \textbf{\bibinfo{volume}{107}}, \bibinfo{pages}{065301}
  (\bibinfo{year}{2011}).

\bibitem[{\citenamefont{Zhang et~al.}(2012)\citenamefont{Zhang, Wang, and
  Zhu}}]{Zhang2012}
\bibinfo{author}{\bibfnamefont{D.-w.} \bibnamefont{Zhang}},
  \bibinfo{author}{\bibfnamefont{Z.-d.} \bibnamefont{Wang}}, \bibnamefont{and}
  \bibinfo{author}{\bibfnamefont{S.-l.} \bibnamefont{Zhu}},
  \bibinfo{journal}{Front. Phys.} \textbf{\bibinfo{volume}{7}},
  \bibinfo{pages}{31} (\bibinfo{year}{2012}).

\bibitem[{\citenamefont{Polini et~al.}({2013})\citenamefont{Polini, Guinea,
  Lewenstein, Manoharan, and Pellegrini}}]{Polini2013}
\bibinfo{author}{\bibfnamefont{M.}~\bibnamefont{Polini}},
  \bibinfo{author}{\bibfnamefont{F.}~\bibnamefont{Guinea}},
  \bibinfo{author}{\bibfnamefont{M.}~\bibnamefont{Lewenstein}},
  \bibinfo{author}{\bibfnamefont{H.~C.} \bibnamefont{Manoharan}},
  \bibnamefont{and}
  \bibinfo{author}{\bibfnamefont{V.}~\bibnamefont{Pellegrini}},
  \bibinfo{journal}{{Nat. Nanotechnol.}} \textbf{\bibinfo{volume}{{8}}},
  \bibinfo{pages}{{625}} (\bibinfo{year}{{2013}}).

\bibitem[{\citenamefont{Eckardt et~al.}(2005)\citenamefont{Eckardt, Weiss, and
  Holthaus}}]{Eckardt2005}
\bibinfo{author}{\bibfnamefont{A.}~\bibnamefont{Eckardt}},
  \bibinfo{author}{\bibfnamefont{C.}~\bibnamefont{Weiss}}, \bibnamefont{and}
  \bibinfo{author}{\bibfnamefont{M.}~\bibnamefont{Holthaus}},
  \bibinfo{journal}{Phys. Rev. Lett.} \textbf{\bibinfo{volume}{95}},
  \bibinfo{pages}{260404} (\bibinfo{year}{2005}).

\bibitem[{\citenamefont{Lignier et~al.}(2007)\citenamefont{Lignier, Sias,
  Ciampini, Singh, Zenesini, Morsch, and Arimondo}}]{Lignier2007}
\bibinfo{author}{\bibfnamefont{H.}~\bibnamefont{Lignier}},
  \bibinfo{author}{\bibfnamefont{C.}~\bibnamefont{Sias}},
  \bibinfo{author}{\bibfnamefont{D.}~\bibnamefont{Ciampini}},
  \bibinfo{author}{\bibfnamefont{Y.}~\bibnamefont{Singh}},
  \bibinfo{author}{\bibfnamefont{A.}~\bibnamefont{Zenesini}},
  \bibinfo{author}{\bibfnamefont{O.}~\bibnamefont{Morsch}}, \bibnamefont{and}
  \bibinfo{author}{\bibfnamefont{E.}~\bibnamefont{Arimondo}},
  \bibinfo{journal}{Phys. Rev. Lett.} \textbf{\bibinfo{volume}{99}},
  \bibinfo{pages}{220403} (\bibinfo{year}{2007}).

\bibitem[{\citenamefont{Struck et~al.}({2011})\citenamefont{Struck,
  \"Olschl\"ager, Le~Targat, Soltan-Panahi, Eckardt, Lewenstein, Windpassinger,
  and Sengstock}}]{Struck2011}
\bibinfo{author}{\bibfnamefont{J.}~\bibnamefont{Struck}},
  \bibinfo{author}{\bibfnamefont{C.}~\bibnamefont{\"Olschl\"ager}},
  \bibinfo{author}{\bibfnamefont{R.}~\bibnamefont{Le~Targat}},
  \bibinfo{author}{\bibfnamefont{P.}~\bibnamefont{Soltan-Panahi}},
  \bibinfo{author}{\bibfnamefont{A.}~\bibnamefont{Eckardt}},
  \bibinfo{author}{\bibfnamefont{M.}~\bibnamefont{Lewenstein}},
  \bibinfo{author}{\bibfnamefont{P.}~\bibnamefont{Windpassinger}},
  \bibnamefont{and}
  \bibinfo{author}{\bibfnamefont{K.}~\bibnamefont{Sengstock}},
  \bibinfo{journal}{Science} \textbf{\bibinfo{volume}{{333}}},
  \bibinfo{pages}{{996}} (\bibinfo{year}{{2011}}).

\bibitem[{\citenamefont{Struck et~al.}(2012)\citenamefont{Struck,
  \"Olschl\"ager, Weinberg, Hauke, Simonet, Eckardt, Lewenstein, Sengstock, and
  Windpassinger}}]{Struck2012}
\bibinfo{author}{\bibfnamefont{J.}~\bibnamefont{Struck}},
  \bibinfo{author}{\bibfnamefont{C.}~\bibnamefont{\"Olschl\"ager}},
  \bibinfo{author}{\bibfnamefont{M.}~\bibnamefont{Weinberg}},
  \bibinfo{author}{\bibfnamefont{P.}~\bibnamefont{Hauke}},
  \bibinfo{author}{\bibfnamefont{J.}~\bibnamefont{Simonet}},
  \bibinfo{author}{\bibfnamefont{A.}~\bibnamefont{Eckardt}},
  \bibinfo{author}{\bibfnamefont{M.}~\bibnamefont{Lewenstein}},
  \bibinfo{author}{\bibfnamefont{K.}~\bibnamefont{Sengstock}},
  \bibnamefont{and}
  \bibinfo{author}{\bibfnamefont{P.}~\bibnamefont{Windpassinger}},
  \bibinfo{journal}{Phys. Rev. Lett.} \textbf{\bibinfo{volume}{108}},
  \bibinfo{pages}{225304} (\bibinfo{year}{2012}).

\bibitem[{\citenamefont{Hauke et~al.}(2012)\citenamefont{Hauke, Tieleman, Celi,
  \"Olschl\"ager, Simonet, Struck, Weinberg, Windpassinger, Sengstock,
  Lewenstein et~al.}}]{Hauke2012}
\bibinfo{author}{\bibfnamefont{P.}~\bibnamefont{Hauke}},
  \bibinfo{author}{\bibfnamefont{O.}~\bibnamefont{Tieleman}},
  \bibinfo{author}{\bibfnamefont{A.}~\bibnamefont{Celi}},
  \bibinfo{author}{\bibfnamefont{C.}~\bibnamefont{\"Olschl\"ager}},
  \bibinfo{author}{\bibfnamefont{J.}~\bibnamefont{Simonet}},
  \bibinfo{author}{\bibfnamefont{J.}~\bibnamefont{Struck}},
  \bibinfo{author}{\bibfnamefont{M.}~\bibnamefont{Weinberg}},
  \bibinfo{author}{\bibfnamefont{P.}~\bibnamefont{Windpassinger}},
  \bibinfo{author}{\bibfnamefont{K.}~\bibnamefont{Sengstock}},
  \bibinfo{author}{\bibfnamefont{M.}~\bibnamefont{Lewenstein}},
  \bibnamefont{et~al.}, \bibinfo{journal}{Phys. Rev. Lett.}
  \textbf{\bibinfo{volume}{109}}, \bibinfo{pages}{145301}
  (\bibinfo{year}{2012}).

\bibitem[{\citenamefont{Roth and Burnett}(2003)}]{Roth2003}
\bibinfo{author}{\bibfnamefont{R.}~\bibnamefont{Roth}} \bibnamefont{and}
  \bibinfo{author}{\bibfnamefont{K.}~\bibnamefont{Burnett}},
  \bibinfo{journal}{Phys. Rev. A} \textbf{\bibinfo{volume}{68}},
  \bibinfo{pages}{023604} (\bibinfo{year}{2003}).

\bibitem[{\citenamefont{Rabl et~al.}(2003)\citenamefont{Rabl, Daley, Fedichev,
  Cirac, and Zoller}}]{Rabl2003}
\bibinfo{author}{\bibfnamefont{P.}~\bibnamefont{Rabl}},
  \bibinfo{author}{\bibfnamefont{A.~J.} \bibnamefont{Daley}},
  \bibinfo{author}{\bibfnamefont{P.~O.} \bibnamefont{Fedichev}},
  \bibinfo{author}{\bibfnamefont{J.~I.} \bibnamefont{Cirac}}, \bibnamefont{and}
  \bibinfo{author}{\bibfnamefont{P.}~\bibnamefont{Zoller}},
  \bibinfo{journal}{Phys. Rev. Lett.} \textbf{\bibinfo{volume}{91}},
  \bibinfo{pages}{110403} (\bibinfo{year}{2003}).

\bibitem[{\citenamefont{Buonsante and Vezzani}(2004)}]{Buonsante2004a}
\bibinfo{author}{\bibfnamefont{P.}~\bibnamefont{Buonsante}} \bibnamefont{and}
  \bibinfo{author}{\bibfnamefont{A.}~\bibnamefont{Vezzani}},
  \bibinfo{journal}{Phys. Rev. A} \textbf{\bibinfo{volume}{70}},
  \bibinfo{pages}{033608} (\bibinfo{year}{2004}).

\bibitem[{\citenamefont{Rousseau et~al.}(2006)\citenamefont{Rousseau, Arovas,
  Rigol, H\'ebert, Batrouni, and Scalettar}}]{Rousseau2006}
\bibinfo{author}{\bibfnamefont{V.~G.} \bibnamefont{Rousseau}},
  \bibinfo{author}{\bibfnamefont{D.~P.} \bibnamefont{Arovas}},
  \bibinfo{author}{\bibfnamefont{M.}~\bibnamefont{Rigol}},
  \bibinfo{author}{\bibfnamefont{F.}~\bibnamefont{H\'ebert}},
  \bibinfo{author}{\bibfnamefont{G.~G.} \bibnamefont{Batrouni}},
  \bibnamefont{and} \bibinfo{author}{\bibfnamefont{R.~T.}
  \bibnamefont{Scalettar}}, \bibinfo{journal}{Phys. Rev. B}
  \textbf{\bibinfo{volume}{73}}, \bibinfo{pages}{174516}
  (\bibinfo{year}{2006}).

\bibitem[{\citenamefont{Chen et~al.}(2010)\citenamefont{Chen, Kou, Zhang, and
  Chen}}]{Chen2010}
\bibinfo{author}{\bibfnamefont{B.-L.} \bibnamefont{Chen}},
  \bibinfo{author}{\bibfnamefont{S.-P.} \bibnamefont{Kou}},
  \bibinfo{author}{\bibfnamefont{Y.}~\bibnamefont{Zhang}}, \bibnamefont{and}
  \bibinfo{author}{\bibfnamefont{S.}~\bibnamefont{Chen}},
  \bibinfo{journal}{Phys. Rev. A} \textbf{\bibinfo{volume}{81}},
  \bibinfo{pages}{053608} (\bibinfo{year}{2010}).

\bibitem[{\citenamefont{Buonsante
  et~al.}(2004{\natexlab{a}})\citenamefont{Buonsante, Penna, and
  Vezzani}}]{Buonsante2004b}
\bibinfo{author}{\bibfnamefont{P.}~\bibnamefont{Buonsante}},
  \bibinfo{author}{\bibfnamefont{V.}~\bibnamefont{Penna}}, \bibnamefont{and}
  \bibinfo{author}{\bibfnamefont{A.}~\bibnamefont{Vezzani}},
  \bibinfo{journal}{Phys. Rev. A} \textbf{\bibinfo{volume}{70}},
  \bibinfo{pages}{061603} (\bibinfo{year}{2004}{\natexlab{a}}).

\bibitem[{\citenamefont{Buonsante and Vezzani}(2005)}]{Buonsante2005a}
\bibinfo{author}{\bibfnamefont{P.}~\bibnamefont{Buonsante}} \bibnamefont{and}
  \bibinfo{author}{\bibfnamefont{A.}~\bibnamefont{Vezzani}},
  \bibinfo{journal}{Phys. Rev. A} \textbf{\bibinfo{volume}{72}},
  \bibinfo{pages}{013614} (\bibinfo{year}{2005}).

\bibitem[{\citenamefont{Buonsante et~al.}(2005)\citenamefont{Buonsante, Penna,
  and Vezzani}}]{Buonsante2005b}
\bibinfo{author}{\bibfnamefont{P.}~\bibnamefont{Buonsante}},
  \bibinfo{author}{\bibfnamefont{V.}~\bibnamefont{Penna}}, \bibnamefont{and}
  \bibinfo{author}{\bibfnamefont{A.}~\bibnamefont{Vezzani}},
  \bibinfo{journal}{Phys. Rev. A} \textbf{\bibinfo{volume}{72}},
  \bibinfo{pages}{031602} (\bibinfo{year}{2005}).

\bibitem[{\citenamefont{Danshita et~al.}(2007)\citenamefont{Danshita, Williams,
  S\'a~de Melo, and Clark}}]{Danshita2007}
\bibinfo{author}{\bibfnamefont{I.}~\bibnamefont{Danshita}},
  \bibinfo{author}{\bibfnamefont{J.~E.} \bibnamefont{Williams}},
  \bibinfo{author}{\bibfnamefont{C.~A.~R.} \bibnamefont{S\'a~de Melo}},
  \bibnamefont{and} \bibinfo{author}{\bibfnamefont{C.~W.} \bibnamefont{Clark}},
  \bibinfo{journal}{Phys. Rev. A} \textbf{\bibinfo{volume}{76}},
  \bibinfo{pages}{043606} (\bibinfo{year}{2007}).

\bibitem[{\citenamefont{Muth et~al.}(2008)\citenamefont{Muth, Mering, and
  Fleischhauer}}]{Muth2008}
\bibinfo{author}{\bibfnamefont{D.}~\bibnamefont{Muth}},
  \bibinfo{author}{\bibfnamefont{A.}~\bibnamefont{Mering}}, \bibnamefont{and}
  \bibinfo{author}{\bibfnamefont{M.}~\bibnamefont{Fleischhauer}},
  \bibinfo{journal}{Phys. Rev. A} \textbf{\bibinfo{volume}{77}},
  \bibinfo{pages}{043618} (\bibinfo{year}{2008}).

\bibitem[{\citenamefont{L\"uhmann}(2013)}]{Luhmann2013}
\bibinfo{author}{\bibfnamefont{D.-S.} \bibnamefont{L\"uhmann}},
  \bibinfo{journal}{Phys. Rev. A} \textbf{\bibinfo{volume}{87}},
  \bibinfo{pages}{043619} (\bibinfo{year}{2013}).

\bibitem[{\citenamefont{Konabe et~al.}(2006)\citenamefont{Konabe, Nikuni, and
  Nakamura}}]{Konabe2006}
\bibinfo{author}{\bibfnamefont{S.}~\bibnamefont{Konabe}},
  \bibinfo{author}{\bibfnamefont{T.}~\bibnamefont{Nikuni}}, \bibnamefont{and}
  \bibinfo{author}{\bibfnamefont{M.}~\bibnamefont{Nakamura}},
  \bibinfo{journal}{Phys. Rev. A} \textbf{\bibinfo{volume}{73}},
  \bibinfo{pages}{033621} (\bibinfo{year}{2006}).

\bibitem[{\citenamefont{Pisarski et~al.}(2011)\citenamefont{Pisarski, Jones,
  and Gooding}}]{Pisarski2011}
\bibinfo{author}{\bibfnamefont{P.}~\bibnamefont{Pisarski}},
  \bibinfo{author}{\bibfnamefont{R.~M.} \bibnamefont{Jones}}, \bibnamefont{and}
  \bibinfo{author}{\bibfnamefont{R.~J.} \bibnamefont{Gooding}},
  \bibinfo{journal}{Phys. Rev. A} \textbf{\bibinfo{volume}{83}},
  \bibinfo{pages}{053608} (\bibinfo{year}{2011}).

\bibitem[{\citenamefont{Freericks and Monien}(1994)}]{Freericks1994}
\bibinfo{author}{\bibfnamefont{J.}~\bibnamefont{Freericks}} \bibnamefont{and}
  \bibinfo{author}{\bibfnamefont{H.}~\bibnamefont{Monien}},
  \bibinfo{journal}{Eur. Phys. Lett.} \textbf{\bibinfo{volume}{26}},
  \bibinfo{pages}{545} (\bibinfo{year}{1994}).

\bibitem[{\citenamefont{Freericks and Monien}(1996)}]{Freericks1996}
\bibinfo{author}{\bibfnamefont{J.}~\bibnamefont{Freericks}} \bibnamefont{and}
  \bibinfo{author}{\bibfnamefont{H.}~\bibnamefont{Monien}},
  \bibinfo{journal}{Phys. Rev. B} \textbf{\bibinfo{volume}{53}},
  \bibinfo{pages}{2691} (\bibinfo{year}{1996}).

\bibitem[{\citenamefont{Buonsante
  et~al.}(2004{\natexlab{b}})\citenamefont{Buonsante, Penna, and
  Vezzani}}]{Buonsante2004c}
\bibinfo{author}{\bibfnamefont{P.}~\bibnamefont{Buonsante}},
  \bibinfo{author}{\bibfnamefont{V.}~\bibnamefont{Penna}}, \bibnamefont{and}
  \bibinfo{author}{\bibfnamefont{A.}~\bibnamefont{Vezzani}},
  \bibinfo{journal}{Phys. Rev. B} \textbf{\bibinfo{volume}{70}},
  \bibinfo{pages}{184520} (\bibinfo{year}{2004}{\natexlab{b}}).

\bibitem[{\citenamefont{Jain and Gardiner}(2004)}]{Jain2004}
\bibinfo{author}{\bibfnamefont{P.}~\bibnamefont{Jain}} \bibnamefont{and}
  \bibinfo{author}{\bibfnamefont{C.~W.} \bibnamefont{Gardiner}},
  \bibinfo{journal}{J. Phys. B: At. Mol. Opt. Phys.}
  \textbf{\bibinfo{volume}{37}}, \bibinfo{pages}{3649} (\bibinfo{year}{2004}).

\bibitem[{\citenamefont{Hen and Rigol}(2009)}]{Hen2009}
\bibinfo{author}{\bibfnamefont{I.}~\bibnamefont{Hen}} \bibnamefont{and}
  \bibinfo{author}{\bibfnamefont{M.}~\bibnamefont{Rigol}},
  \bibinfo{journal}{Phys. Rev. B} \textbf{\bibinfo{volume}{80}},
  \bibinfo{pages}{134508} (\bibinfo{year}{2009}).

\bibitem[{\citenamefont{McIntosh et~al.}(2012)\citenamefont{McIntosh, Pisarski,
  Gooding, and Zaremba}}]{McIntosh2012}
\bibinfo{author}{\bibfnamefont{T.}~\bibnamefont{McIntosh}},
  \bibinfo{author}{\bibfnamefont{P.}~\bibnamefont{Pisarski}},
  \bibinfo{author}{\bibfnamefont{R.~J.} \bibnamefont{Gooding}},
  \bibnamefont{and} \bibinfo{author}{\bibfnamefont{E.}~\bibnamefont{Zaremba}},
  \bibinfo{journal}{Phys. Rev. A} \textbf{\bibinfo{volume}{86}},
  \bibinfo{pages}{013623} (\bibinfo{year}{2012}).

\bibitem[{\citenamefont{Yamamoto et~al.}(2012)\citenamefont{Yamamoto, Masaki,
  and Danshita}}]{Yamamoto2012b}
\bibinfo{author}{\bibfnamefont{D.}~\bibnamefont{Yamamoto}},
  \bibinfo{author}{\bibfnamefont{A.}~\bibnamefont{Masaki}}, \bibnamefont{and}
  \bibinfo{author}{\bibfnamefont{I.}~\bibnamefont{Danshita}},
  \bibinfo{journal}{Phys. Rev. B} \textbf{\bibinfo{volume}{86}},
  \bibinfo{pages}{054516} (\bibinfo{year}{2012}).

\bibitem[{\citenamefont{Gawryluk et~al.}(2013)\citenamefont{Gawryluk,
  Miniatura, and Gr{\'e}maud}}]{Gawryluk2013}
\bibinfo{author}{\bibfnamefont{K.}~\bibnamefont{Gawryluk}},
  \bibinfo{author}{\bibfnamefont{C.}~\bibnamefont{Miniatura}},
  \bibnamefont{and}
  \bibinfo{author}{\bibfnamefont{B.}~\bibnamefont{Gr{\'e}maud}},
  \bibinfo{journal}{arXiv:1212.4570v2}  (\bibinfo{year}{2013}).

\end{thebibliography}
\end{document}